\documentclass[journal]{IEEEtran}

\usepackage{amsmath,amsfonts}
\usepackage{float}
\usepackage{algorithmic}
\usepackage{algorithm}
\usepackage{siunitx}
\usepackage[normalem]{ulem}
\usepackage{dsfont}
\usepackage{mleftright}
\usepackage{bbm}

\usepackage{tikz}
\usetikzlibrary{spy, arrows.meta,arrows}

\begin{document}

\title{List-based Optimization of Proximal Decoding for LDPC Codes }

\author{Andreas Tsouchlos, Holger Jäkel, and Laurent Schmalen, \IEEEmembership{Fellow, IEEE}
\thanks{The authors are with the Communications Engineering Lab (CEL), Karlsruhe Institute of Technology (KIT), corresponding author: \texttt{holger.jaekel@kit.edu}
This work has received funding in part from the European Research Council
(ERC) under the European Union’s Horizon 2020 research and innovation
program (grant agreement No. 101001899).
\\
© 2024 IEEE. Personal use of this material is permitted. Permission from
IEEE must be obtained for all other uses, in any current or future media,
including reprinting/republishing this material for advertising or promotional
purposes, creating new collective works, for resale or redistribution to servers
or lists, or reuse of any copyrighted component of this work in other works.
Article DOI: : 10.1109/LCOMM.2024.3458422
}
 }

\markboth{Submitted to IEEE Communications Letters}{List-based Optimization of Proximal Decoding for Linear Block Codes}

\maketitle

%
%

\begin{abstract}
In this paper, the proximal decoding algorithm is considered within the
context of \textit{additive white Gaussian noise} (AWGN) channels.
An analysis of the convergence behavior of the algorithm shows that
proximal decoding inherently enters an oscillating behavior of the estimate
after a certain number of iterations.
Due to this oscillation, frame errors arising during decoding can often
be attributed to only a few remaining wrongly decoded bit positions.
In this letter, an improvement of the proximal decoding algorithm is proposed
by establishing an additional step, in which these erroneous positions are
attempted to be corrected.
We suggest an empirical rule with which the components most likely needing
correction can be determined.
Using this insight and performing a subsequent ``ML-in-the-list'' decoding,
a gain of up to 1 dB is achieved compared to conventional
proximal decoding, depending on the decoder parameters and the code.
\end{abstract}


%
%

\section{Introduction}

\IEEEPARstart{C}{hannel} coding using binary linear codes is a way of enhancing
the reliability of data by detecting and correcting any errors that may occur
during its transmission or storage.
One class of binary linear codes, \textit{low-density parity-check} (LDPC)
codes, has become especially popular due to its ability to reach arbitrarily
small error probabilities at code rates up to the capacity of the channel
\cite{mackay99}, while retaining a structure that allows for very efficient
decoding.
While the established decoders for LDPC codes, such as belief propagation (BP)
and the min-sum algorithm, offer good decoding performance, they are generally
not optimal and exhibit an error floor for high
\textit{signal-to-noise ratios} (SNRs) \cite{channel_codes_book}, rendering them
inadequate for applications with extreme reliability requirements.

Optimization based decoding algorithms are an entirely different way of
approaching the decoding problem:
they map the decoding problem onto an optimization problem in order to
leverage the vast knowledge from the field of optimization theory.
A number of different such algorithms have been introduced in the literature.
The field of \textit{linear programming} (LP) decoding \cite{feldman_paper},
for example, represents one class of such algorithms, based on a relaxation
of the \textit{maximum likelihood} (ML) decoding problem as a linear program.
Many different optimization algorithms can be used to solve the resulting
problem \cite{ADMM, adaptive_lp_decoding, interior_point_decoding}.
Recently, proximal decoding for LDPC codes was presented by
Wadayama \textit{et al.} \cite{proximal_paper}.
Proximal decoding relies on a non-convex optimization formulation
of the \textit{maximum a posteriori} (MAP) decoding problem.

The aim of this work is to improve the performance of proximal decoding by
first presenting an analysis of the algorithm's behavior and then suggesting
an approach to mitigate some of its flaws.
This analysis is performed for
\textit{additive white Gaussian noise} (AWGN) channels.
We first observe that the algorithm initially moves the estimate in
the right direction; however, in the final steps of the decoding process,
convergence to the correct codeword is often not achieved.
Subsequently, we attribute this behavior to the nature
of the decoding algorithm itself, comprising two separate gradient descent
steps working adversarially.

We, thus, propose a method to mitigate this effect by appending an
additional step to the iterative decoding process.
In this additional step, the components of the estimate with the highest
probability of being erroneous are identified.
New codewords are then generated, over which an ``ML-in-the-list''
\cite{ml_in_the_list} decoding is performed.
The main point of the paper at hand is to improve list generation such that
it is especially tailored to the nature of proximal decoding. 
Using the improved algorithm, a gain of up to
$\SI{1}{dB}$ can be achieved compared to conventional proximal decoding,
depending on the decoder parameters and the code.

\section{Preliminaries}

\subsection{Notation}

When considering binary linear codes, data words are mapped onto
codewords, the lengths of which are denoted by $k \in \mathbb{N}$
and $n \in \mathbb{N}$, respectively, with $k \le n$.
The set of codewords $\mathcal{C} \subset \mathbb{F}_2^n$ of a binary linear
code can be characterized using the parity-check matrix
$\boldsymbol{H} \in \mathbb{F}_2^{m \times n} $, where $m$ represents the
number of parity-checks:
\begin{align*}
    \mathcal{C} := \left\{ \boldsymbol{c} \in \mathbb{F}_2^n :
        \boldsymbol{H}\boldsymbol{c}^\text{T} = \boldsymbol{0} \right\}
\end{align*}

The check nodes indexed by $j \in \mathcal{J}:=\left\{1, \ldots, m\right\}$ 
correspond to the parity checks, i.e., to the rows of $\boldsymbol{H}$.
The variable nodes indexed by $i \in \mathcal{I}:=\left\{1, \ldots, n\right\}$ correspond
to the components of a codeword, i.e., to the columns of $\boldsymbol{H}$.
The neighborhood of a parity check $j$, i.e., the set of component indices
relevant for the according parity check, is denoted by
$\mathcal{N}_\mathrm{c}(j) := \left\{i \in \mathcal{I}: \boldsymbol{H}\negthinspace_{j,i} = 1 \right\},
\hspace{2mm} j \in \mathcal{J}$.

In order to transmit a codeword $\boldsymbol{c} \in \mathbb{F}_2^n$, it is
mapped onto a \textit{binary phase shift keying} (BPSK) symbol via
$\boldsymbol{x} = 1 - 2\boldsymbol{c}$, with
$ \boldsymbol{x} \in \left\{\pm 1\right\}^n$, which is then transmitted over an
AWGN channel.
The received vector $\boldsymbol{y} \in \mathbb{R}^n$ is decoded to obtain an
estimate $\hat{\boldsymbol{c}} \in \mathbb{F}_2^n$ of the transmitted codeword.
A distinction is made between $\boldsymbol{x} \in \left\{\pm 1\right\}^n$
and $\tilde{\boldsymbol{x}} \in \mathbb{R}^n$,
the former denoting the transmitted BPSK symbols and
the latter being used as a variable during the optimization process.
The likelihood of receiving $\boldsymbol{y}$ upon transmitting
$\boldsymbol{x}$ is expressed by the \textit{probability density function} (PDF)
$f_{\boldsymbol{Y}\mid\boldsymbol{X}}(\boldsymbol{y} \mid \boldsymbol{x})$.

\subsection{Proximal Decoding}
With proximal decoding, the proximal gradient method \cite{proximal_algorithms}
is used to solve a non-convex optimization formulation of the MAP decoding
problem.
With the equal prior probability assumption for all codewords, MAP and ML
decoding are equivalent and, specifically for AWGN channels, correspond to a
nearest-neighbor decision.
For this reason, decoding can be carried out using a figure of merit that
describes the distance from a given vector to a codeword.
One such expression, formulated under the assumption of BPSK, is the
\textit{code-constraint polynomial} defined in \cite{proximal_paper}
\begin{align*}
    h( \tilde{\boldsymbol{x}} ) =
        \underbrace{\sum_{i=1}^{n}
            \left( \tilde{x}_i^2-1 \right) ^2}_{\text{Bipolar constraint}}
        + \underbrace{\sum_{j=1}^{m} \left[
            \left( \prod_{i\in \mathcal{N}_\mathrm{c} \left( j \right) } \tilde{x}_i \right)
        -1 \right] ^2}_{\text{Parity constraint}}
.\end{align*}%
Its intent is to penalize vectors far from a codeword and, thus, it serves as an objective function describing the quality of possible
estimates.
Please note that all valid codewords are local minima of $h(\tilde{\boldsymbol{x}})$. 
The code-constraint polynomial comprises two terms: the first part is
representing the bipolar constraint due to using BPSK, whereas the second part
is representing the parity constraint, incorporating all information regarding
the code. 
Please note that the first part of the code-constraint polynomial
may be easily adapted to higher order constellations, whereas the second part of
the code-constraint polynomial requires bit values in $\mathbbm{F}_2$.
This can be achieved by employing a bit-metric decoder.

The channel can be characterized using the negative log-likelihood
$
	L \mleft( \boldsymbol{y} \mid \tilde{\boldsymbol{x}} \mright) = -\ln (
    f_{\boldsymbol{Y} \mid \tilde{\boldsymbol{X}}} \mleft(
	    \boldsymbol{y} \mid \tilde{\boldsymbol{x}} \mright))
$
.
Then, the information about the channel and the code are consolidated in the
objective function \cite{proximal_paper}
\begin{align*}
    g \mleft( \tilde{\boldsymbol{x}} \mright)
        = L \mleft( \boldsymbol{y} \mid \tilde{\boldsymbol{x}} \mright)
            + \gamma h\mleft( \tilde{\boldsymbol{x}} \mright),
        \hspace{5mm} \gamma > 0%
.\end{align*}
The objective function $g \mleft( \tilde{\boldsymbol{x}} \mright)$ is minimized
using the proximal gradient method, which amounts to iteratively performing two
gradient-descent steps \cite{proximal_paper} with the given objective function
in AWGN channels.
To this end, two helper variables $\boldsymbol{r}$ and $\boldsymbol{s}$ are
introduced, describing the result of each of the two steps:
\begin{alignat}{3}
    \boldsymbol{r} &\leftarrow \boldsymbol{s}
        - \omega \nabla L\mleft( \boldsymbol{y} \mid \boldsymbol{s} \mright),
        \hspace{5mm }&&\omega > 0, \label{eq:r_update}\\
    \boldsymbol{s} &\leftarrow \boldsymbol{r}
        - \gamma \nabla h\mleft( \boldsymbol{r} \mright),
        \hspace{5mm} &&\gamma > 0 \label{eq:s_update}
.\end{alignat}
Derivation of
$\nabla  L\mleft( \boldsymbol{y} \mid \boldsymbol{s} \mright) = \boldsymbol{s} - \boldsymbol{y}$
for AWGN and an equation for determining $\nabla h(\boldsymbol{r})$ are given
in \cite{proximal_paper}, where it is also proposed to initialize
$\boldsymbol{s}=\boldsymbol{0}$.
It should be noted that $\boldsymbol{r}$ and $\boldsymbol{s}$ represent
$\tilde{\boldsymbol{x}}$ during different stages of the decoding process.

As the gradient of the code-constraint polynomial can attain very large values
in some cases, an additional step is introduced in \cite{proximal_paper} to
ensure numerical stability:
every estimate $\boldsymbol{s}$ is projected onto the hypercube
$\left[-\eta, \eta\right]^n$ by a projection
$\Pi_\eta : \mathbb{R}^n \rightarrow \left[-\eta, \eta\right]^n$ 
defined as component-wise clipping, i.e.,
$\Pi_\eta(x_i) = \arg \min_{-\eta\leq \xi \leq \eta } |x_i-\xi|$
as in \cite{proximal_paper}, 
where $\eta$ is a positive constant larger than one, e.g., $\eta = 1.5$.
The resulting decoding process is given in Algorithm \ref{alg:proximal_decoding}.

\begin{algorithm}
	\caption{Proximal decoding in AWGN \cite{proximal_paper}}
    \label{alg:proximal_decoding}

    \begin{algorithmic}[1]
        \STATE $\boldsymbol{s} \leftarrow \boldsymbol{0}$
        \STATE \textbf{for} $K$ iterations \textbf{do}
        \STATE \hspace{5mm} $\boldsymbol{r} \leftarrow \boldsymbol{s} - \omega \left( \boldsymbol{s} - \boldsymbol{y} \right) $
        \STATE \hspace{5mm} $\boldsymbol{s} \leftarrow \Pi_\eta \left(\boldsymbol{r} - \gamma \nabla h\left( \boldsymbol{r} \right) \right)$
		\STATE \hspace{5mm} $\boldsymbol{\hat{c}} \leftarrow \mathbbm{1}_{\left\{ \boldsymbol{s} \le 0 \right\}}$
        \STATE \hspace{5mm} \textbf{if} $\boldsymbol{H}\boldsymbol{\hat{c}} = \boldsymbol{0}$ \textbf{do}
        \STATE \hspace{10mm} \textbf{return} $\boldsymbol{\hat{c}}$
        \STATE \textbf{return} $\boldsymbol{\hat{c}}$
    \end{algorithmic}
\end{algorithm}

\section{Improved algorithm}

\subsection{Analysis of the Convergence Behavior}

\begin{figure}[htb]
    \centering

 	\includegraphics{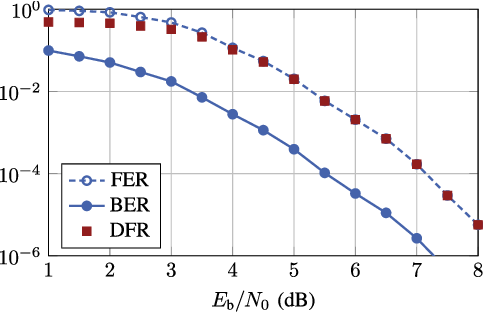}

    \caption{FER, DFR, and BER for $\left( 3, 6 \right)$-regular LDPC code with
        $n=204, k=102$ \cite[\text{204.33.484}]{mackay}.
        Parameters: 
        $\gamma =0.05,\omega = 0.05,
        \eta = 1.5, K=200$.
    }
    \label{fig:fer vs ber}
\end{figure}

In Fig. \ref{fig:fer vs ber}, the \textit{frame error rate} (FER),
\textit{bit error rate} (BER), and \textit{decoding failure rate} (DFR) of
proximal decoding are shown for the LDPC code [204.33.484] \cite{mackay} with
$n=204$ and $k=102$.
Hereby, a \emph{decoding failure} is defined as returning a
\emph{non valid codeword}, i.e., as non-convergence of the algorithm.
The parameters chosen in this simulation are $\gamma=0.05$, $\omega=0.05$,
$\eta=1.5$, and $K=200$ ($K$ describing the maximum number of iterations).
They adhere to \cite{proximal_paper} and 
were determined to offer the best performance in a preliminary examination, where
the effect of changing multiple parameters was simulated over a wide range of
values.
It is apparent that the DFR completely dominates the FER for sufficiently high
SNR.
This means that most frame errors are not due to the algorithm converging to the
wrong codeword, but due to the algorithm not converging at all.

As proximal decoding is an optimization-based decoding method, one possible
explanation for this effect might be that during the decoding process, convergence
to the final codeword is often not achieved, although the estimate is moving
into the right direction.
This would suggest that most frame errors occur due to only a few incorrectly
decoded bits.

An approach for lowering the FER might then be to add an ``ML-in-the-list''
\cite{ml_in_the_list} step to the decoding process shown in Algorithm
\ref{alg:proximal_decoding}.
This step consists in determining the $N \in \mathbb{N}$ positions
$\mathcal{I}'\subset \mathcal{I}$ of bits that are most probably erroneous,
generating a list of $2^N$ codeword candidates out of the current estimate
$\hat{\boldsymbol{c}}$ with bits in $\mathcal{I}'$ adopting all possible values,i.e., 
\begin{equation}\label{eq:def:L_prime}
    \mathcal{L}'=\left\{ \hat{\boldsymbol{c}}'\in\mathbb{F}_2^n: \hat{c}'_i=\hat{c}_i, i\notin \mathcal{I}'\text{ and } \hat{c}'_i\in\mathbb{F}_2, i\in \mathcal{I}'  \right\},
\end{equation}
and performing ML decoding on this list. 
Whereas list decoding is usually based on the analysis of received
values, e.g., ML-in-the-list decoding or Chase decoding \cite{chase_decoding},
the following consideration aims at generating this list by exploiting
characteristic properties of proximal decoding.

\begin{figure}[htb]
    \centering

 	\includegraphics{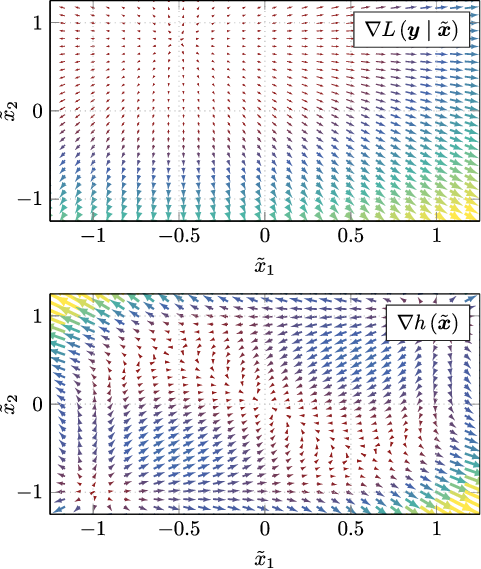}

    \caption{Gradients
        $\nabla L\left(\boldsymbol{y} \mid \tilde{\boldsymbol{x}}\right)$
        and $\nabla h \left( \tilde{\boldsymbol{x}} \right)$ for a repetition
        code with $n=2$.
        Shown for $\boldsymbol{y} = \begin{pmatrix} -0.5 & 0.8 \end{pmatrix}$.
    }
    \label{fig:grad}
\end{figure}

The aforementioned process crucially relies on identifying the positions of bits
that are most likely  erroneous.
Therefore, the convergence properties of proximal decoding are investigated.
Fig. \ref{fig:grad} shows the two gradients performed for a repetition code with
$n=2$.
It is apparent that a net movement will result as long as the two gradients have
a common component.
As soon as this common component is exhausted, they will work in opposing
directions resulting in an oscillation of the estimate.
This behavior supports the conjecture that the reason for the high DFR is a
failure to converge to the correct codeword in the final steps of the
optimization process.

\begin{figure}[htb]
    \centering

 	\includegraphics{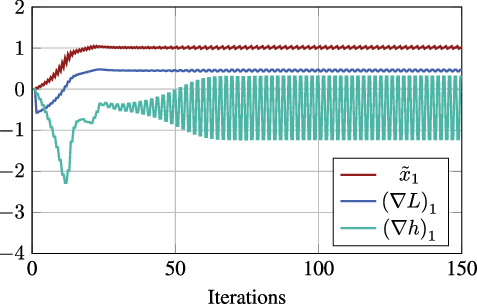}

    \caption{Visualization of component $\tilde{x}_1$
        for a decoding operation for a (3,6) regular LDPC code with
        $n=204, k=102$ \cite[\text{204.33.484}]{mackay}.
        Parameters:
        $\gamma = 0.05, \omega = 0.05,
        \eta = 1.5, E_b/N_0 = \SI{6}{dB}$.
    }
    \label{fig:prox:convergence_large_n}
\end{figure}

In Fig. \ref{fig:prox:convergence_large_n}, we show the component $\tilde{x}_1$
and corresponding gradients during  decoding for the [204.33.484] LDPC code.
We observe that both gradients start oscillating after a certain number of
iterations.
Furthermore, it can be observed the both gradients have approximately equal
average magnitudes, but possess opposing signs, leading to an oscillation of
$\tilde{x}_1$.

\subsection{Improvement Using ``ML-in-the-List'' Step}

\begin{figure}[hbt]
    \centering

	 \includegraphics{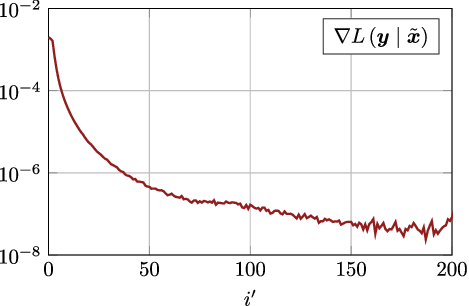}

    \caption{Probability that $P(\hat{c}_{i'} \ne c_{i'})$ for a (3,6) regular
        LDPC code with $n=204, k=102$ \cite[\text{204.33.484}]{mackay}.
        Indices $i'$ are ordered as in eq. (\ref{eq:def:i_prime}).
        Parameters: $\gamma = 0.05, \omega = 0.05, \eta = 1.5,
        E_\text{b}/N_0 = \SI{6}{dB}$,
        $10^9$ codewords.
    }
    \label{fig:p_error}
\end{figure}

Based on the observations depicted in Fig. \ref{fig:grad} and Fig.
\ref{fig:prox:convergence_large_n}, it seems a meaningful approach to tag the
$N\in\mathbb{N}$ most likely erroneous bits based on the oscillation of the
gradient of the code-constraint polynomial.
To this end, let
$\Delta_i^{(h)}:=| \left(\nabla h\right)_{i}[K] - \left(\nabla h\right)_{i}[K-1]|$
be the oscillation height at the last iteration with $\left(\nabla h\right)_{i}[K]$
denoting the gradient at position $i$ and iteration $K$.
Now, let $\boldsymbol{i}'=(i'_1, \ldots, i_n')$ be a permutation of
$\left\{1, \ldots, n\right\}$ such that $\Delta_{i'}^{(h)}$ is arranged according
to increasing oscillation height and select its $N$ smallest indices, i.e.,
\begin{align}\label{eq:def:i_prime}
    & \boldsymbol{i}'=(i_1',\ldots, i_n')\in S_n : 
    \left| \Delta_{i_1'}^{(h)}\right| 
    \leq\cdots\leq
    \left| \Delta_{i_n'}^{(h)} \right|
    \\       
    \label{eq:def:set_I}
    &\mathcal{I}'=\{i_1', \ldots, i_N'\} \text{ with } \boldsymbol{i}' \text{ as defined in (\ref{eq:def:i_prime})}
\end{align}
with $S_n$ denoting the symmetric group of $\{1,\ldots, n\}$.
To reason this approach, Fig. \ref{fig:p_error} shows Monte Carlo simulations of
the probability that the decoded bit $\hat{c}_{i'}$ at position $i'$ of the estimated codeword is wrong.
It can be observed that lower magnitudes of oscillation height correlate with a
higher probability that the corresponding bit was not decoded correctly.
Thus, the oscillation height might be used as a feasible indicator for
identifying the $N$ bits that are most likely erroneous.

The proposed improved algorithm is given in Algorithm \ref{alg:improved}.
First, the proximal decoding Algorithm \ref{alg:proximal_decoding} is applied.
If a valid codeword has been reached, i.e., if the algorithm has converged,
we return this solution.
Otherwise, $N \in \mathbb{N}$ components are selected  as described in eq. (\ref{eq:def:set_I}).
Originating from $\boldsymbol{\hat{c}} \in \mathbb{F}_2^n$, the result of proximal decoding,
the list $\mathcal{L}'$ of codeword candidates with bits in $\mathcal{I}'$ modified is
generated and an ``ML-in-the-list'' step is performed. 
If the list $\mathcal{L}'$ does not contain a valid codeword and, thus,
$\mathcal{L}'_\text{valid}=\emptyset$, the additional step boils down to the
maximization of $2^N$ correlations
$\langle 1-2\boldsymbol{c}_l', \boldsymbol{y}\rangle $, $\boldsymbol{c}_l'\in\mathcal{L}'$,
in which ties happen with probability zero and are solved arbitrarily.
Note that $2^N$ parity checks have to be evaluated for elements in $\mathcal{L}'$
in order to determine $\mathcal{L}'_\text{valid}$ either way.
Restricting the correlations to the (non-empty) list $\mathcal{L}'_\text{valid}$
may reduce the computational burden and ensure that a valid codeword is returned.

\begin{algorithm}
    \caption{Proposed improved proximal decoding in AWGN }
    \label{alg:improved}

    \begin{algorithmic}[1]

        \STATE \text{$\hat{\boldsymbol{c}}\leftarrow\text{proximal decoding}(\boldsymbol{y})$}
        
        \STATE \textbf{if} $\boldsymbol{H}\boldsymbol{\hat{c}} = \boldsymbol{0}$ \textbf{do}
        \STATE \hspace{5mm} \textbf{return} $\boldsymbol{\hat{c}}$
        \STATE $\textcolor{black}{\text{$\mathcal{I}'\leftarrow \{i_1',\ldots, i_N'\}$ (indices of $N$ probably wrong bits) } }$

        \STATE $\textcolor{black}{\text{ 
        $\mathcal{L}'\leftarrow\left\{ \boldsymbol{\hat{c}}'\in\mathbb{F}_2^n: \hat{c}'_i=\hat{c}_i, i\notin \mathcal{I}' \text{ and } \hat{c}'_i\in\mathbb{F}_2, i\in \mathcal{I}'  \right\}$ 
        }}
        $\vspace{1mm}

        \STATE \textcolor{black}{$\mathcal{L}'_\text{valid} \leftarrow \{ \boldsymbol{\hat{c}}'\in\mathcal{L}': \boldsymbol{H}\boldsymbol{\hat{c}}'=\boldsymbol{0}\}$ (select valid codewords) }
        \STATE \textcolor{black}{\textbf{if} $\mathcal{L}'_\text{valid}\neq\emptyset$ \textbf{do}}
        \STATE \hspace{5mm} 
        \textcolor{black}{\textbf{return} $\arg\max \{ \langle 1-2\boldsymbol{\hat{c}}'_l, \boldsymbol{y} \rangle : \boldsymbol{\hat{c}}'_l\in\mathcal{L}'_\text{valid}\}$}
        \STATE \textcolor{black}{\textbf{else}}
        \STATE \hspace{5mm} 
        \textcolor{black}{\textbf{return} $\arg\max \{ \langle 1-2 \boldsymbol{\hat{c}}'_l, \boldsymbol{y} \rangle : \boldsymbol{\hat{c}}'_l\in\mathcal{L}'\}$}
    \end{algorithmic}
\end{algorithm}

\section{Simulation Results \& Discussion}

\begin{figure}[hbt]
    \centering

 	\includegraphics{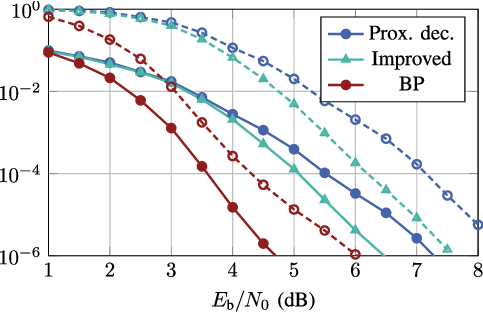}

    \caption{FER (-\,-\,-) and BER (---) of proximal decoding \cite{proximal_paper} and the
        improved algorithm for a $\left( 3, 6 \right)$-regular LDPC code with
        $n=204, k=102$ \cite[\text{204.33.484}]{mackay}.
        Parameters:
        $\gamma=0.05, \omega=0.05, \eta=1.5,
        K=200, N=8$.}

    \label{fig:results}
\end{figure}

Fig. \ref{fig:results} shows the FER and BER resulting from applying
proximal decoding as presented in \cite{proximal_paper} and the proposed improved
algorithm, when both are applied to the $\left( 3,6 \right)$-regular LDPC
code [204.33.484] \cite{mackay} with $n=204$ and $k=102$.
The parameters chosen for the simulation are $\gamma = 0.05, \omega=0.05, \eta=1.5, K=200$ 
as for proximal decoding, since those parameters also turned out close-to-optimum
for the improved algorithm in our simulations. 
The number of possibly wrong components was selected as $N=8$. 
To reason this choice, Table \ref{N Table} shows the SNRs required for $N\in\{4, 6, 8, 10, 12\}$ to achieve an FER of $10^{-2}$ and $10^{-3}$, respectively.

\begin{table}[hbt]
    \centering
   \caption{SNR (in dB) to achieve target FERs $10^{-2}$ and $10^{-3}$}
    \label{N Table}
    \includegraphics{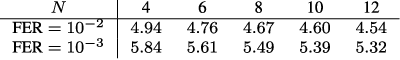}
\end{table}

A noticeable improvement can be observed both in the FER and the BER.
The gain varies significantly with the SNR, which is to be expected since higher
SNR values result in a decreased number of bit errors, making the correction of
those errors in the ``ML-in-the-list'' step more likely.
For an FER of $10^{-6}$, the gain is approximately $\SI{1}{dB}$. 
As shown in Fig. \ref{fig:results}, it can be seen that BP decoding 
with $200$ iterations outperforms the improved scheme by approximately $\SI{1.7}{dB}$.
Nevertheless, note that Algorithm \ref{alg:improved} requires only linear
operations and could be favorable in  applications as, e.g., massive MIMO, in
which application of BP is prohibitive \cite{proximal_paper}. 
Similar behavior to Fig. \ref{fig:results} was observed with a number of
different codes, e.g., \cite[\text{PEGReg252x504, 204.55.187, 96.3.965}]{mackay}.
Furthermore, we did not observe an immediate relationship between the code length
and the gain during our examinations.

\section{Conclusion}

In this paper, an improvement on proximal decoding as presented by
Wadayama \emph{et al.} \cite{proximal_paper} is proposed for AWGN channels.
It relies on the fact that most errors observed in proximal decoding stem
from only a few components of the estimate being wrong.
These few erroneous components can mostly be corrected by appending an
additional step to the original algorithm that is only executed if the
algorithm has not converged.
A gain of up to $\SI{1}{dB}$ can be observed, depending on the code,
the code parameters, and the SNR.

%


\begin{thebibliography}{10}
	\providecommand{\url}[1]{#1}
	\csname url@samestyle\endcsname
	\providecommand{\newblock}{\relax}
	\providecommand{\bibinfo}[2]{#2}
	\providecommand{\BIBentrySTDinterwordspacing}{\spaceskip=0pt\relax}
	\providecommand{\BIBentryALTinterwordstretchfactor}{4}
	\providecommand{\BIBentryALTinterwordspacing}{\spaceskip=\fontdimen2\font plus
		\BIBentryALTinterwordstretchfactor\fontdimen3\font minus
		\fontdimen4\font\relax}
	\providecommand{\BIBforeignlanguage}[2]{{%
			\expandafter\ifx\csname l@#1\endcsname\relax
			\typeout{** WARNING: IEEEtran.bst: No hyphenation pattern has been}%
			\typeout{** loaded for the language `#1'. Using the pattern for}%
			\typeout{** the default language instead.}%
			\else
			\language=\csname l@#1\endcsname
			\fi
			#2}}
	\providecommand{\BIBdecl}{\relax}
	\BIBdecl
	
	\bibitem{mackay99}
	D.~MacKay, ``Good error-correcting codes based on very sparse matrices,''
	\emph{IEEE Trans. Inf. Theory}, vol.~45, no.~2, pp. 399--431, 1999.
	
	\bibitem{channel_codes_book}
	W.~Ryan and S.~Lin, \emph{Channel Codes: Classical and Modern}.\hskip 1em plus
	0.5em minus 0.4em\relax Cambridge University Press, 2009.
	
	\bibitem{feldman_paper}
	J.~Feldman, M.~Wainwright, and D.~Karger, ``Using linear programming to decode
	binary linear codes,'' \emph{IEEE Trans. Inf. Theory}, vol.~51, no.~3, pp.
	954--972, 2005.
	
	\bibitem{ADMM}
	S.~Barman, X.~Liu, S.~C. Draper, and B.~Recht, ``Decomposition methods for
	large scale lp decoding,'' \emph{Trans. Inf. Theory}, vol.~59, no.~12, pp.
	7870--7886, 2013.
	
	\bibitem{adaptive_lp_decoding}
	M.~H. Taghavi and P.~H. Siegel, ``Adaptive linear programming decoding,'' in
	\emph{IEEE Proc. ISIT}, 2006, pp. 1374--1378.
	
	\bibitem{interior_point_decoding}
	P.~O. Vontobel, ``Interior-point algorithms for linear-programming decoding,''
	in \emph{Proc. ITA}, 2008, pp. 433--437.
	
	\bibitem{proximal_paper}
	T.~Wadayama and S.~Takabe, ``Proximal decoding for {LDPC} codes,'' \emph{IEICE
		Transactions on Fundamentals of Electronics, Communications and Computer
		Sciences}, 2022.
	
	\bibitem{ml_in_the_list}
	M.~Geiselhart, A.~Elkelesh, M.~Ebada, S.~Cammerer, and S.~t. Brink,
	``Automorphism ensemble decoding of reed–muller codes,'' \emph{IEEE Trans.
		Commun.}, vol.~69, no.~10, pp. 6424--6438, 2021.
	
	\bibitem{proximal_algorithms}
	N.~Parikh, S.~Boyd \emph{et~al.}, \emph{Proximal algorithms}.\hskip 1em plus
	0.5em minus 0.4em\relax Now Publishers, Inc., 2014, vol.~1, no.~3.
	
	\bibitem{mackay}
	\BIBentryALTinterwordspacing
	D.~J. MacKay. Encyclopedia of sparse graph codes. [Online]. Available:
	\url{www.inference.org.uk/mackay/codes/data.html}
	\BIBentrySTDinterwordspacing
	
	\bibitem{chase_decoding}
	D.~Chase, ``Class of algorithms for decoding block codes with channel
	measurement information,'' \emph{IEEE Trans. Inf. Theory}, 1972.
	
\end{thebibliography}
\end{document}